\documentclass[prd,11pt,nofootinbib,preprintnumbers,final]{revtex4}
\usepackage{graphicx}
\usepackage{amssymb}
\usepackage{amsmath}
\usepackage{srcltx}
\usepackage{ulem}

\renewcommand{\(}{\left(}
\renewcommand{\)}{\right)}
\renewcommand{\[}{\left[}
\renewcommand{\]}{\right]}
\newcommand{\nn}{\nonumber}

\begin{document}
\preprint{LU TP 14-01} \preprint{RUB-TPII-05/2013}

\title{Key features of the TMD soft-factor structure}

\author{A.~A.~Vladimirov}
\affiliation{Department of Astronomy and Theoretical Physics, Lund University,\\
           S\"olvegatan 14A, S 223 62 Lund, Sweden}
\email{vladimirov.aleksey@gmail.com}

\author{N.~G.~Stefanis}
\affiliation{Institut f\"ur Theoretische Physik II, Ruhr-
           Universit\"at Bochum, 44780 Bochum, Germany }
\email{stefanis@tp2.ruhr-uni-bochum.de}

\begin{abstract}
We show that the geometry of the Wilson lines, entering the operator definition of the transverse-momentum dependent parton distributions and
that of the soft factor, follows from the kinematics of the underlying physical process in conjunction with the gauge invariance of the QCD
Lagrangian. We demonstrate our method in terms of concrete examples and determine the paths of the associated Wilson lines. The validation of
the factorization theorem in our approach is postponed to future work.

\end{abstract}

\maketitle

\section{Introduction}
\label{sec:intro} Parton distribution functions, integrated and those
that retain the partonic transverse degrees of freedom unintegrated,
have to include Wilson lines (gauge links) in order to render their
definitions gauge invariant (see, \cite{Boer:2011fh} for a recent
review).
There are several ways to obtain the Wilson-line structure of the
parton distribution operator in deeply inelastic scattering (DIS).
The most known procedure is the direct resummation of the collinear
gluon radiation diagrams, see, e.g.,
\cite{Collins:1989gx,Collins:2011zzd}.
This method can be applied to transverse momentum dependent (TMD)
processes as well, and it allows to determine the appropriate
configuration of Wilson lines in the parton distribution functions
(PDFs) that parameterize them \cite{Collins:1989gx,Collins:2011zzd}.
On the other hand, the geometry of the soft factor does not directly
follow from the resummation procedure, but is rather related to the
removal of rapidity singularities
\cite{Collins:2011ca,Cherednikov:2007tw,Cherednikov:2008ua,Cherednikov:2009wk}.
A counterexample is the approach used in \cite{GarciaEchevarria:2011rb},
based on the soft collinear effective theory (SCET), in which one can
establish a TMD factorization theorem and explicitly derive the soft
factor with the appropriate Wilson lines.
The drawback of the SCET approach is that one cannot really pass back
to the original QCD fields, and, hence, a direct comparison of the
operator expressions is questionable.

In this note we want to discuss a procedure which allows one to reveal the geometry of the Wilson lines in hard processes. Although the
presented method cannot replace the exact proof of the TMD factorization theorem without additional specifications, it can serve as a
demonstration of the universality and uniqueness of the soft-factor definition. The main idea is to use the freedom of the gauge redefinition of
the fields in the QCD Lagrangian and the fact that the choice of a particular gauge can be implemented by performing an explicit gauge rotation
(that will depend on the adopted configuration of the gluon fields). This gauge rotation allows us to freely pass from an operator defined in a
particular gauge to a manifestly gauge-invariant definition. In conjunction with the kinematic constraints, it enables us to use the power of
the SCET approach within pure QCD to pick up the correct Wilson-line geometry. For the sake of clarity, we start with a heuristic discussion to
demonstrate this idea in the case of DIS, and consider then the TMD factorization of the Drell-Yan (DY) process as an application of this
technique. The details of the full-fledged analysis will be given in \cite{SV14-in-preparation}.

\section{Educational example: DIS}
\label{sec:DIS}
The hadron tensor for the DIS process reads
\begin{eqnarray}
  W^{\mu\nu}
=
  \frac{(2\pi)^2}{2}\int d^4\xi\;
  e^{iq\cdot\xi}\;
  \langle p|T\left(J^\mu(\xi)J^\nu(0)\right)|p\rangle \, ,
\end{eqnarray}
where $q$ has the large virtuality $-q^2=Q^2\gg\Lambda^2_{\text{QCD}}$, and $J^{\mu}=\bar q \gamma^\mu q$ is the electromagnetic current. It is
well known (see, e.g., \cite{Collins:1989gx}) that the proof of the factorization theorem drastically simplifies if one uses a special reference
frame, notably, the infinite-momentum frame (with the spatial part of the hadron momentum taken along the third axis with a large component
$p_+\to \infty$) and imposing the lightcone gauge $A_+=0$. Here and below, we use the standard notation for the lightcone components $v^\mu=\bar
n^\mu v^++n^\mu  v^-+v_\perp^\mu$, with $n^2=\bar n^2=0$ and $(n\cdot\bar n)=1$. The operator for the parton distribution in the lightcone gauge
is just a bilocal quark operator: $:\!\!\bar q(\xi)q(0)\!\!:$, where $\xi$ lies on the light-cone. Note that we use the following four-vector
notation: $\xi^\mu = (\xi^+, \xi^-, \bm{\xi}_\perp)$. For simplicity, we do not display these indices in the figures below.

The general --- gauge independent --- expression is harder to derive.
The derivation requires the complete resummation of the collinear
gluon vertices --- see e.g., \cite{Collins:2011zzd} --- and amounts
to the insertion of a gauge link between the quarks depicting a contour
that is compatible with the physical situation under study.
This is done for a particular version of the eikonal approximation ---
termed the soft approximation \cite{Collins:1981uk}.
In that case, the gauge contour can be taken to be a light ray.
Note that from the renormalization-group point of view, the particular
contour of the gauge link is of minor importance as long as it is
smooth, because only its endpoints create additional divergences that
can be removed by dimensional regularization and minimal subtraction
\cite{CS80,Ste84,Ste12_LC2012}.
Contours with cusps and/or self-intersections deserve special care
\cite{CS80,Aoy81}.
The TMD case is more complicated and the proper Wilson line has to be
defined with prudence \cite{Belitsky:2002sm}.
In fact, it was first shown in
\cite{Cherednikov:2007tw,Cherednikov:2008ua,Cherednikov:2009wk},
and later also in other works \cite{GarciaEchevarria:2011rb}, that
the ultraviolet (UV) divergences mix with the rapidity divergences that
ensue from cusps in the Wilson line which joins two points with a
finite transverse separation.

However, one can obtain the gauge-independent expression by a much
simpler method.
The essence of this method is the following:
Knowing the operator in the lightcone gauge, one can perform a gauge
transformation of the quark fields from the lightcone gauge to a form
that is manifestly gauge invariant.
Indeed, for every configuration of the gluon field, there exists a
particular gauge transformation which fixes one or another gauge
condition.
The gauge transformation which neglects the ``plus''-component of the
gluon field, and hence satisfies the light-cone gauge, can be found
from the equation
\begin{eqnarray}
  U A_+(x) U^\dagger-\frac{1}{ig}U \partial_+ U^\dagger=0 \, .
\label{UAU=UdU}
\end{eqnarray}
Because the lightcone gauge is not exhausting the gauge freedom, we should also fix the boundary conditions imposed on the gluon propagator and
adopt, say, the retarded boundary condition, $A_+(x)\big|_{x^-\to -\infty^-}=0$. We denote the solution of equation (\ref{UAU=UdU}) by $W_+$:
\begin{eqnarray}
  W_+[A(x)]
=
  \[x,-\infty^+|\bar{n}^\mu\]
=
  \text{Pexp}\(-ig\int^0_{-\infty}~ d\sigma
  A_+(x^\mu+\bar{n}^\mu\sigma)\) \, ,
\label{W_def}
\end{eqnarray}
where the gauge field $A^\mu$ here and below is a shorthand notation
for $\sum_{a}t^a A_{a}^\mu$.
Expression (\ref{W_def}) describes a Wilson line along the direction
$\bar{n}^\mu$ connecting the point $x$ with infinity, where the gluon
field is supposed to vanish by virtue of the imposed boundary condition.

\begin{figure}
\centering
  \includegraphics[width=0.6\textwidth]{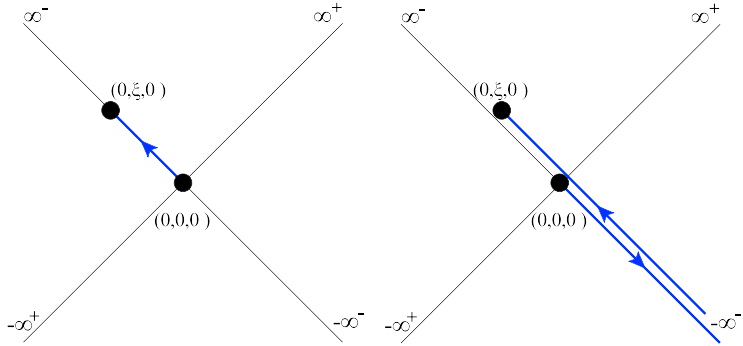}
\caption{Illustration of the gauge-invariance restoration in the integrated parton operator for DIS via gauge links. The left panel shows the
direct contour $\Gamma$, while the right one depicts the contour through infinity along the segments $\Gamma_1$ and $\Gamma_2$ that corresponds
to the gauge transformation $W_-$.} \label{fig:1}
\end{figure}

In this way, the operator for the integrated parton distribution
obtained in the lightcone gauge can be written in gauge-invariant form
with the help of the gauge transformation in terms of the matrix $W_+$,
i.e.,
\begin{eqnarray}
  \bar q\(\xi\)q\(0\)
~\to~
  \bar q\(\xi\)W_+^\dagger\(\xi\)W_+\(0\) q\(\xi\)
=
  \bar q\(\xi\)\[\xi,0\Big|n^\mu\]q\(0\),~~(\xi^+=\xi_\perp=0) \, .
\label{DIS_res}
\end{eqnarray}
This result (illustrated in Fig.\ \ref{fig:1}) is well known and it
can be obtained in a number of ways \cite{Ste12_LC2012}.
It expresses the restoration of the geometry of the Wilson line,
starting from an expression that contains no Wilson lines at all.

The important point here is that this treatment picks up the route
of the Wilson line through infinity and not the direct one --- see
Fig.\ \ref{fig:1}).
The latter would appear as the difference of two successive $SU(3)_c$
rotations first from $\xi^-$ to $-\infty^-$ along $n$, followed by a
rotation from $0^-$ to $-\infty^-$.
Would one use another boundary condition, it would be possible to
direct the Wilson line to $\infty^-$.
A possible non-triviality might appear if one would impose a boundary
condition that would allow the gluon field to be different from zero at
lightcone infinity.
Then, in order to satisfy Eq.\ (\ref{UAU=UdU}), one would have to
multiply the solution by an additional boundary-dependent factor, viz.,
the Wilson line from the lightcone infinity to the point, where the
gluon field vanishes.
This factor would not depend on $A_+$.
The simplest solution would be to add a gauge link in the transverse
direction, which can also be obtained by direct resummation as in
\cite{Belitsky:2002sm}, or be deduced from a SCET analysis like in
\cite{GarciaEchevarria:2011md}.

\section{Drell-Yan process, and SCET}
\label{sec:DY-SCET}
Let us now consider processes that depend on unintegrated PDFs.
The main difficulty in the description of such processes is the
intensive exchange of soft gluons between the two initial (final) hadrons.
These exchanges lead to a double logarithmic asymptotic behavior and
constitute the main perturbative contribution of the factorized hard part.
Factorization reorganizes the singularity structure of the initial
amplitude, leaving a trace in terms of the anomalous dimensions of the
operators.
In the case of the TMD factorization, the leading singularities
amount to double logarithms.
To collect and reorganize these singularities, an operator is necessary,
termed the soft factor.
It is composed of Wilson lines which effectively accumulate all soft-gluon
exchanges.
Presumably, different configurations of Wilson lines may give rise to the
same singularity structure in fixed-order perturbation theory.
On the other hand, the initial object --- the hadron tensor --- does not
contain any singularities so that it is up to us to define the soft factor
and the parton density operators in a self-consistent way.

Let us now test our calculational scheme by applying it to the Drell-Yan
process.
The standard way to factorize this process can be found in
\cite{Collins:1984kg,Collins:1989gx,Collins:2011zzd}.
A perhaps simpler way to derive a factorized expression employs SCET
\cite{GarciaEchevarria:2011rb}.
Recently, it was shown that these two approaches yield equivalent results
\cite{Collins:2012uy}, albeit the
regularization of rapidity divergences is treated differently.
The virtue of the SCET approach is that the Wilson-line geometry of the
soft factor and that of the parton-distribution operators follows from
the very structure of the SCET electromagnetic currents.
In contrast, in standard QCD the construction of the Wilson-line
geometry is, presumably, not uniquely defined.
Indeed, within the standard approach the soft
factor is defined empirically with the aim to render the final
expressions free of rapidity divergences.
Consistency checks of such procedures have not yet been established
beyond the one-loop level.

In order to show that the geometry of the Wilson lines within the TMD
factorization follows from the kinematics of the process, we will first
rewrite the dynamics in factorized form using SCET.
Subsequently, we will pass back to the original field degrees of
freedom of QCD.
This procedure will resemble the derivation of the Wilson-line structure
in the DIS case, considered in the previous section.
To this end, we are going to use for expedience the lightcone gauge and
the so-called hadron frame \cite{Collins:1981uk} together with SCET-like
variables and then express the fields in a general gauge using an
$SU(3)_c$ rotation.
The formal proof of factorization within this calculational scheme is
beyond the scope of this short exposition.
Here, we only want to show the uniqueness of the gauge-link geometry of
the hard processes obtained this way.

In the hadron frame the two hadrons, involved in the DY scattering,
move nearly on the light-cone in opposite directions: hadron
$A$ along $\bar n^\mu$ and hadron $B$ along $n^\mu$.
The large lightcone components of the hadron's momentum suggest the
following decomposition of the spinor fields into ``large'' and ``small''
components \cite{Kogut:1969xa}, indicated by $+$ and $-$ labels,
respectively:
\begin{eqnarray}
  q(x)
=
  Q_+(x)+Q_-(x),~~~~~Q_\pm(x)
=
  \frac{\gamma^\pm\gamma^\mp}{2}q(x) \, .
\label{eq:spin-dec}
\end{eqnarray}
It is straightforward to show that the $Q_+$($Q_-$) component of the
field propagates mainly along the $\bar n$($n$) direction, and,
therefore, according to the equation of motion, the hadron $A$ ($B$)
consists mainly of the $Q_+$($Q_-$) components of the quark fields.
The contribution of the other components ($Q_{-}$($Q_{+}$) in hadron
$A$($B$)) is suppressed by
$M_{A}/P^+_A\sim M_{B}/P^-_B\sim \mathcal{O}(1/Q)$,
see, e.g., \cite{Kogut:1969xa}.
Thus, in leading order of $Q^2$, we can neglect the mixing of the
quark components inside the hadron.

Then, the quark sector of the QCD Lagrangian in terms of the $Q_\pm$
components reads
\begin{eqnarray}
  \mathcal{L}^q_{QCD}
=
   \bar Q_- \gamma^+D_- Q_-
  +\bar Q_+ \gamma^-D_+ Q_+
  +\bar Q_+ \gamma^\mu_\perp D^\perp_\mu Q_-
  +\bar Q_- \gamma^\mu_\perp D^\perp_\mu Q_+ \, .
\label{eq:Q+-Lag}
\end{eqnarray}
We consider the fields $Q_\pm$ as being independent non-mixing fields,
whereas the mixing terms ($\sim \bar Q_\pm Q_\mp$) can be viewed as a
kind of ``contact interaction''.
This consideration is natural due to the fact that $Q_+$ talks to $Q_-$
only by means of the transverse derivative.
And since these fields carry the large plus (minus) components of the
available momentum and the transverse components are small, their
admixtures can be treated as small corrections $\sim p_\perp/Q$.
[Note in passing that the transverse components of the momenta are
small because their only source is the transverse motion of the partons
inside the hadron $\sim \Lambda_{\text{QCD}}$ and the external source
$k_T^2\ll Q^2$.
Note also that the loop-momentum does not bear a large transverse
component --- see, \cite{SV14-in-preparation} for details.]
Thus, in the hadron frame, QCD effectively splits into two
``one-dimensional'' theories (see, for instance, \cite{Bassetto:1996ph}
and references cited therein).
The term ``one-dimensional'' reflects the fact that the fields $Q_\pm$
move only along the light-rays $p_\pm$

This situation may look deceptively simple but in reality it is more complicated because of gluon exchanges between the quark fields. In fact,
the gluons can mix the components of the $Q$ fields without $k_T$ suppression, as we know from the application of factorization theorems
\cite{Collins:2011zzd}. This demands the resummation of all diagrams with soft-gluon radiations to the final state, a cumbersome procedure owing
to the $k_T$ dependence and the interaction between the two initial hadrons in the DY process. The discussion could be significantly simplified
were we able to pass to a gauge, in which the gluon radiation is suppressed --- such as in the DIS case. But there is no single gauge condition
that can prevent the different components of the quark fields from emitting (absorbing) different components of the gluon fields. However, since
we have two ``one-dimensional'' theories, we can fix the gauge for each ``theory'' separately with the help of the gauge rotation matrix
$W_\pm$. (The matrix $W_{+}[A(x)]=[x,-\infty^+|\bar{n}^\mu]$ was defined before, whereas $W_{-}$ corresponds to the Wilson line
$W_{-}[A(0)]=[0,-\infty^{-} |n^{\mu}]$.) We transform the corresponding components $Q$ individually via the matrices $W$ (assuming for
simplicity that $A_\mu(-\infty)=0$)). The Lagrangian for the transformed fields reads
\begin{eqnarray}
  \mathcal{L}^q_{QCD}
& = &
    \bar Q^W_- \gamma^+\partial_- Q^W_-
    +\bar Q^W_+ \gamma^-\partial_+ Q^W_+
    +\bar Q^W_+ W_+^\dagger \gamma^\mu_\perp D^\perp_\mu W_- Q^W_-
    +\bar Q^W_- W_-^\dagger \gamma^\mu_\perp D^\perp_\mu W_+ Q^W_+ \, ,
\label{L_qw}
\end{eqnarray}
where
$
 Q^W_\pm(x)
=
 W^\dagger_\pm(x)\gamma^\pm\gamma^\mp q(x)/2$.
The Lagrangian (\ref{L_qw}) is nonlocal because the last two terms
contain the interaction of the quarks with the gluons in the Wilson
lines.
However, this nonlocality is not a real problem, since these
contributions are suppressed by the transverse momentum components.
Hence, the nonlocal vertices represent the suppressed part of the
gluon radiation (the so-called, ultrasoft gluons in the SCET
terminology), whereas the unsuppressed collinear gluons are hidden
inside the fields $Q^W$.

\begin{figure}
\centering
  \includegraphics[width=0.8\textwidth]{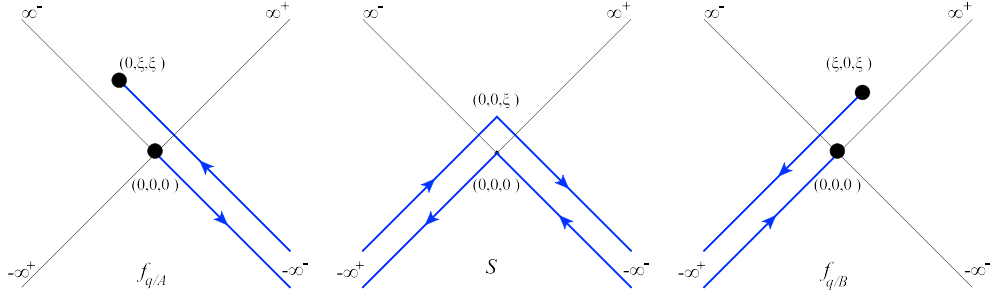}
\caption{Contours of the Wilson lines for the components of the DY
cross section: quark operator associated with hadron A (left),
soft factor (center), and quark operator pertaining to hadron B
(right).
Assembling together the Wilson lines, most parts of them cancel
--- see text.}
\label{fig:2}       
\end{figure}

Since the cross-talk interactions of the fields $Q_W$ are suppressed by
$k_T/Q$ and the non-mixing terms belong to two ``one-dimensional'' free
theories, the derivation of the factorized form for the hadronic tensor
is straightforward and can be expressed as a matrix element of a bilocal
current operator to read
\begin{eqnarray}
  W^{\mu\nu}
=
  s\int d^4 \xi ~e^{iq\cdot\xi}~
  \langle p_A,p_B|J^\mu(\xi)J^\nu(0)|p_A,p_B\rangle \, .
\label{W_DY}
\end{eqnarray}
The product of the currents can be cast in the form
\begin{eqnarray}
\label{JJ}
  J^\mu(\xi)J_\mu(0)
=
    (\bar Q^W_-\gamma^+Q^W_-)(\xi)(\bar Q^W_+\gamma^-Q^W_+)(0)
  + (\bar Q^W_+W^\dagger_+\gamma_\perp^\mu W_-Q^W_-)(\xi)(\bar Q^W_+W^\dagger_+\gamma_\perp^\mu W_-Q^W_-)(0)
\\ \nn
  +...+ (\bar Q^W_+W^\dagger_+\gamma_\perp^\mu W_-Q^W_-)(\xi)(\bar Q^W_-W^\dagger_-\gamma_\perp^\mu W_+Q^W_+)(0) \, ,
\end{eqnarray}
where for brevity we have considered the Lorentz trace of the currents.
One has in total six terms in this expansion.
Because the hadrons are composed in leading $1/Q$-order of different
quark components, the four terms of the current expansion
(e.g., the terms presented in the first line of Eq.\ (\ref{JJ})
are suppressed. The remaining two terms can be factorized according
to the free theory, i.e., by a straightforward Wick contraction of the
fields. As a result, the gluon fields attached to the Wilson lines
appear to be decoupled from the hadron matrix elements.
For example, the last term in (\ref{JJ}) has the following
decomposition
\begin{eqnarray}
&&
  \int d^4\xi e^{-iq\cdot\xi}\langle N_1,N_2|(\bar Q^W_+W^\dagger_+\gamma_\perp^\mu W_-Q^W_-)(\xi)
  (\bar Q^W_-W^\dagger_-\gamma_\perp^\mu W_+Q^W_+)(0)|N_1N_2\rangle
  \\\nn
&=&
  \int d^4\xi e^{-iq\cdot\xi}\langle N_1|\bar Q^{a,i}_-(\xi)Q^{b,j}_-(0)|N_1\rangle\langle N_2|\bar
  Q^{c,k}_+(\xi)Q^{d,l}_+(0)|N_2\rangle \langle [W_-^+W_+]_{ad}(0)[W_+^\dagger
  W_-]_{cb}(\xi)\rangle\gamma^{\mu_\perp}_{il}\gamma^{\mu_\perp}_{kj} \, ,
\label{JJ-last-term}
\end{eqnarray}
where the indices $a,b,c,d$ belong to the color group, and $i,j,k,l$
are spinor indices.

The open color and Lorentz indices can be contracted with the help of
Fiertz transformations and the resulting expression contains a large
number of different terms with all possible combinations of Lorentz and
color indices.
However, most operators are averaged to zero due to color conservation
and Lorentz invariance.
Collecting all the terms together, we obtain for the hadron tensor
the following expression
\begin{eqnarray}
  W^{\mu\nu}
& \sim &
  \sum_{f,l}C^{\mu\nu}(p_A,p_B,Q^2)
  \int d^2\bm{\xi}_\perp e^{i(\bm{q}_\perp \cdot \bm{\xi}_\perp)}
  f^l_{f/A}(x_A,\bm{\xi}_\perp)f^l_{f/B}(x_B, \bm{\xi}_\perp)S(\bm{\xi}_\perp) \, ,
\label{W_DY_factor}
\end{eqnarray}
where $f_{f/A}$ represents the TMD PDF of quark $f$ in hadron $A$
(with an analogous definition for hadron $B$) and is given by
\begin{eqnarray}
&& f^l_{f/A}(x,\bm{\xi}_\perp)
=
   \int \frac{d\xi^-}{2\pi}e^{-ix p_A^+\xi^-}
   \langle p_A|\bar Q^W_+(\xi)\gamma^-_l Q^W_+|p_A\rangle\Big|_{\xi^+=0} \, ,
\label{TMD_PDF_DY}
\end{eqnarray}
whereas $S$ is the soft factor
\begin{eqnarray}
  S(\bm{\xi}_\perp)
=
  \frac{1}{N_c}\langle 0|
  \text{Tr}(W_-^\dagger(0)W_+(0)W_+^\dagger(\bm{\xi}_\perp)W_-(\bm{\xi}_\perp))
               |0\rangle \, .
\label{Soft_DY}
\end{eqnarray}
The index $l$ stands for different combinations of $\gamma$ matrices, such as $\mathbf{1}$ and $\gamma^5$. It is worth noting that the soft
factor depends only on the transverse coordinate $\bm{\xi}_\perp$, while the dependence on the lightcone coordinates $\xi^\pm$ is contained only
inside the PDFs. The graphical illustration of the Wilson lines entering the TMD PDFs is given in Fig.\ \ref{fig:2}.

The upshot of the above procedure is that the structure of the soft factor in terms of Wilson lines is uniquely determined  \textit{before}
entering an explicit loop calculation. A loop calculation would unavoidably necessitate the introduction of the soft factor in order to remove
terms containing admixtures of UV and rapidity divergences. Although our presented analysis has not formally proved factorization, we have shown
that the geometry of the Wilson lines is determined by construction --- at least in leading order of $k_T/Q$.

The final step in our procedure is to perform the inverse substitution
$Q^W\to q$ and obtain the expressions for the PDF operators in QCD ---
in complete analogy to the DIS case.
The results coincide with the standard definitions of these operators.
A key issue of this procedure is that the Wilson lines in the PDF
operators reflect the structure of the soft factor.
Thus, taking the product of the operators, the Wilson lines overlap and
mostly cancel each other.
The remaining parts correspond to those Wilson-line segments that are
implied by the (leading-order) factorization procedure.
Exactly the same cancelation takes place in the integrated case.
In the unintegrated case, the half-infinite Wilson lines play an important
role because they accumulate information about the divergences related to
the contour-induced divergences.
These divergences, also known as rapidity divergences, cancel
order-by-order in the perturbative expansion of the product of the soft
factor and the PDFs.

\section{Conclusions}
\label{concl}
We have presented a method to derive the Wilson-line structure of
the soft factor entering the definition of TMD PDFs akin to SCET.
This method has the virtue of being precisely defined via the QCD
Lagrangian and the process kinematics in terms of the hadron tensor.
Therefore, the gauge-link geometry for the soft factor and the TMD
PDF operators is enshrined in its conception and does not have to
be imposed a posteriori.
We discussed the application of this calculational scheme to the
Drell-Yan process.



\begin{thebibliography}{3}
\bibitem{Boer:2011fh}
  Boer, D., et al.:
  Gluons and the quark sea at high energies: Distributions, polarization, tomography.
  arXiv:1108.1713 [nucl-th]

\bibitem{Collins:1989gx}
  Collins, J.C., Soper, D.E., Sterman, G.F.:
  Factorization of hard processes in QCD.
  Adv.\ Ser.\ Direct.\ High Energy Phys.\  {\bf 5}, 1 (1988)

\bibitem{Collins:2011zzd}
  Collins, J.C.:
  Foundations of perturbative QCD.
  Cambridge University Press, Cambridge (2011)

\bibitem{Collins:2011ca}
  Collins, J.C.:
  New definition of TMD parton densities.
  Int.\ J.\ Mod.\ Phys.\ Conf.\ Ser.\  {\bf 4}, 85 (2011)

\bibitem{Cherednikov:2007tw}
  Cherednikov, I.O., Stefanis, N.G.:
  Renormalization, Wilson lines, and transverse-momentum dependent parton distribution functions.
  Phys.\ Rev.\ D {\bf 77}, 094001 (2008)

\bibitem{Cherednikov:2008ua}
  Cherednikov, I.O., Stefanis, N.G.:
  Wilson lines and transverse-momentum dependent parton distribution functions: A renormalization-group analysis.
  Nucl.\ Phys.\ B {\bf 802}, 146 (2008)

\bibitem{Cherednikov:2009wk}
  Cherednikov, I.O., Stefanis, N.G.:
  Renormalization-group properties of transverse-momentum dependent parton distribution functions in the light-cone gauge with the Mandelstam-Leibbrandt prescription.
  Phys.\ Rev.\ D {\bf 80}, 054008 (2009)

\bibitem{GarciaEchevarria:2011rb}
  Echevarria, M.G., Idilbi, A., Scimemi, I.:
  Factorization theorem for Drell-Yan at low $q_T$ and transverse momentum distributions on-the-light-cone.
  JHEP {\bf 1207}, 002 (2012)


\bibitem{SV14-in-preparation} Stefanis, N.G., Vladimirov, A.A., in preparation.

\bibitem{Collins:1981uk}
  Collins, J.C., Soper, D.E.:
  Back-To-Back Jets in QCD.
  Nucl.\ Phys.\ B {\bf 193}, 381 (1981)
  [Erratum: ibid.\ B {\bf 213}, 545 (1983)]


\bibitem{CS80}
  Craigie, N.S., Dorn, H.:
  On the renormalization and short distance properties of hadronic operators in QCD.
  Nucl.\ Phys.\ B {\bf 185}, 204 (1981)

\bibitem{Ste84}
  Stefanis, N.G.:
  Gauge invariant quark two-point Green's function through connector insertion to $O(\alpha_s)$.
  Nuovo Cim.\ A {\bf 83}, 205 (1984)

\bibitem{Ste12_LC2012}
  Stefanis, N.G.:
  Worldline techniques and QCD observables.
  Acta Phys.\ Polon.\ Supp.\ {\bf 6}, 71 (2013)

\bibitem{Aoy81}
  Aoyama, S.:
  The renormalization of the string operator in QCD.
  Nucl.\ Phys.\ B {\bf 194}, 513 (1982)

\bibitem{Belitsky:2002sm}
  Belitsky, A.V., Ji, X., Yuan, F.:
  Final state interactions and gauge invariant parton distributions.
  Nucl.\ Phys.\ B {\bf 656}, 165 (2003)


\bibitem{GarciaEchevarria:2011md}
  Garcia-Echevarria, M., Idilbi, A., Scimemi, I.:
  SCET, light-cone gauge and the T-Wilson lines.
  Phys.\ Rev.\ D {\bf 84}, 011502 (2011)

\bibitem{Collins:1984kg}
  Collins, J.C., Soper, D.E., Sterman, G.F.:
  Transverse momentum distribution in Drell-Yan pair and W and Z boson production.
  Nucl.\ Phys.\ B {\bf 250}, 199 (1985)

\bibitem{Collins:2012uy}
  Collins, J.C., Rogers, T.C.:
  Equality of two definitions for transverse momentum dependent parton distribution functions.
  Phys.\ Rev.\ D {\bf 87}, 034018 (2013)

\bibitem{Kogut:1969xa}
  Kogut, J.B., Soper, D.E.:
  Quantum Electrodynamics in the infinite momentum frame.
  Phys.\ Rev.\ D {\bf 1}, 2901 (1970)

\bibitem{Bassetto:1996ph}
  Bassetto, A.:
  Renormalization in light cone gauge: How to do it in a consistent way.
  Nucl.\ Phys.\ Proc.\ Suppl.\ {\bf 51C}, 281 (1996)
\end{thebibliography}
\end{document}